\documentclass[conference]{IEEEtran}
\IEEEoverridecommandlockouts

\usepackage{cite}
\usepackage{amsmath,amssymb,amsfonts}
\usepackage{graphicx}
\usepackage{textcomp}
\usepackage{xcolor}
\usepackage{url}

\usepackage{array}
\usepackage{multirow}
\usepackage{multicol}
\usepackage{pifont}
\usepackage{booktabs}
\usepackage{xspace}
\usepackage{enumitem}
\usepackage{placeins}
\usepackage{algorithm}
\usepackage{algorithmicx}
\usepackage{algpseudocode}
\usepackage{tikz}
\usepackage[hidelinks]{hyperref}

\renewcommand{\footnoterule}{%
  \kern-3pt
  \hrule width 0.4\columnwidth height 0.4pt
  \kern2.6pt}

\def\BibTeX{{\rm B\kern-.05em{\sc i\kern-.025em b}\kern-.08em
    T\kern-.1667em\lower.7ex\hbox{E}\kern-.125emX}}

\newcommand{\ignore}[1]{}
\newcommand{\mytodo}[1]{}

\newcommand{\comms}[1]{#1}

\newcommand{\mpo}{$<_p$\xspace}
\newcommand{\mvo}{$<_v$\xspace}
\newcommand{\po}{<_p}
\newcommand{\vo}{<_v}

\newcommand{\method}{\textsc{TEMPO}\xspace}
\newcommand{\lfvr}{\textit{got}\xspace}
\newcommand{\vr}{\textit{ot}\xspace}
\newcommand{\vsb}{\textit{vsb}\xspace}
\newcommand{\orq}{\textit{orq}\xspace}

\newcommand{\mynode}[1]
{\tikz\node[circle,scale=0.5,color=white,fill=black]{\large #1};}

\newtheorem{verule}{TEMPO Rule}

\title{TEMPO: A Tag-Based Framework for Efficient Memory Ordering}
\author{\IEEEauthorblockN{Pranith Kumar\textsuperscript{*}}
\IEEEauthorblockA{\textit{Arm}\\
San Jose, USA\\
pranith@gatech.edu}
\and
\IEEEauthorblockN{Prasun Gera\textsuperscript{*}}
\IEEEauthorblockA{\textit{NVIDIA}\\
Santa Clara, USA\\
prasun.gera@gatech.edu}
\and
\IEEEauthorblockN{Hyojong Kim\textsuperscript{*}}
\IEEEauthorblockA{\textit{Samsung Semiconductor Inc.}\\
San Jose, USA\\
hyojong.kim@gatech.edu}
\and[\hfill\break\hspace*{0.20\textwidth}]
\IEEEauthorblockN{Chulhyung Park}
\IEEEauthorblockA{\textit{Georgia Institute of Technology}\\
Atlanta, USA\\
chp@gatech.edu}
\and[\hspace*{0.08\textwidth}]
\IEEEauthorblockN{Hyesoon Kim}
\IEEEauthorblockA{\textit{Georgia Institute of Technology}\\
Atlanta, USA\\
hyesoon@cc.gatech.edu}
\thanks{\textsuperscript{*}This research was done by the authors at
Georgia Institute of Technology.}}

\begin{document}

\maketitle
\begin{abstract}
  Weak-memory processors rely on ordering instructions for correctness, yet
  conventional implementations often enforce them more conservatively than the
  memory model requires. This over-enforcement manifests as drain-induced
  retirement stalls at ordering instructions and conservative squash/replay of
  speculative loads, suppressing legal executions and reducing throughput.

  We present \textit{TEMPO}, a tag-based framework for precise
  microarchitectural implementation of ordering instructions. \method
  assigns lightweight ordering tags to instructions and decomposes
  enforcement across retirement-time predicates and completion-time store
  ordering, allowing the core to enforce required ordering without
  conservative retirement serialization.

  \method eliminates unnecessary retirement serialization at ordering
  instructions and speculative-load squash/replay. In our evaluation,
  \method reduces geometric-mean normalized execution cycles by 7.9\% on
  native four-thread workloads and improves geometric-mean IPC by 15.9\% on
  an instrumented SPEC2017 dynamic binary translation (DBT) proxy for
  cross-ISA execution (e.g., x86-on-Arm), while adding only 262 bytes per core.
\ignore{
  It also
  reduces geometric-mean ordering-instruction ROB residency by 30.8\% (native) and 46.0\%
  (SPEC2017 proxy), and reduces commit stalls waiting for merge-buffer
  drain by 45.7\% on the SPEC2017 proxy.
  }
\end{abstract}


\section{Introduction}

Memory consistency models define the ordering and visibility of memory
operations across threads. Sequential consistency~\cite{lam79,cha:len90}
provides a simple programmer-visible model in which memory operations
appear to execute in a single global order consistent with each thread's
program order. However, this strong ordering restricts reordering and
limits the ability of hardware to overlap memory operations
aggressively. Weak-memory models improve performance by relaxing these
constraints, allowing implementations to exploit buffering, speculation,
and out-of-order execution more effectively. To preserve correctness under
relaxed ordering, modern ISAs pair weak-memory execution with explicit
ordering instructions and synchronization operations~\cite{alg:mar10}.

Release Consistency ($RC$)~\cite{gha:len90,adv:gha96} provides directional
ordering through acquire/release synchronization, rather than imposing
bidirectional ordering on every synchronization event. In this model, a
load-acquire orders only program-order (po)-younger memory operations, while a
store-release orders only po-older memory operations; a full fence, in
contrast, imposes both directions~\cite{adv:gha96,alg:mar10}. Mainstream
weak-memory ISAs expose this same directional structure via acquire/release
operations and explicit fences, including Armv8, PowerPC, and RISC-V
RVWMO~\cite{pul:flu17,mad:mar12,rvwmo}. This asymmetry
creates optimization opportunities that are not available when all
synchronization is treated as full-fence ordering~\cite{liu:zan20}.

Conventional microarchitectures largely ignore this directional
structure. The ordering constraints of full fences and store-release
instructions are commonly enforced by draining older stores before
retirement, which stalls commit, whereas load-acquire instructions may
conservatively squash and replay speculative loads following
invalidations. This conservative handling introduces unnecessary
serialization. Implementations incur both retirement stalls and avoidable
speculation-recovery work, and may preclude executions that are permitted
by the memory model because they do not distinguish order-critical accesses
from accesses that are already safe with respect to in-flight ordering
instructions. This cost is not inherent to the memory model; it is an
artifact of where ordering is enforced.

This problem is increasingly important in weak-memory server and client
platforms. Growing adoption of Arm and RISC-V also increases demand for
efficient x86 binary translation: preserving x86 ordering on these
weak-memory hosts often requires fences~\cite{gou:spr22,gao:men24}.
Fence instructions are fundamental building blocks for
synchronization in shared-memory systems, and synchronization overhead
becomes increasingly sensitive to fence cost as systems scale to larger
core counts~\cite{ber:rie23, gao:fan21}. As a result, conservative ordering
mechanisms impose growing performance costs, increasing the value of precise
ordering enforcement.

Our key insight is to encode ordering constraints as lightweight
per-instruction metadata and enforce them at the relevant pipeline stages,
avoiding unnecessary reorder buffer (ROB) serialization. Based on this
insight, we introduce \textit{TEMPO}, a tag-based memory-ordering framework.
It
encodes ordering state explicitly and enforces only memory-model-required
constraints, thereby avoiding conservative pipeline stalls and unnecessary
speculative-load squash/replay.

Ordering tags are assigned in program order and carried through the
load/store queues and the merge buffer. On the store side, they shift
enforcement from drain-induced retirement stalls to tag-ordered visibility
at merge-buffer completion: stores drain in tag order, so required
visibility constraints are preserved without stalling ROB-head
retirement. On the load side, \method uses tag-aware retirement predicates
to classify loads as \textit{safe} or \textit{unsafe} and defers
invalidation-based hazard handling until the relevant ordering constraint
becomes active. The mechanism is entirely within-core in its base form and
requires no ISA extensions or coherence-protocol changes.

We develop the mechanism in the context of the Release Consistency model
implemented in Armv8~\cite{flu:gra16,pul:flu17}. We use memory fences as
the primary application of this framework; however, the same framework also
applies to other serializing instructions that impose ordering constraints
on surrounding memory operations.


Building on this mechanism, this paper makes three contributions:
\begin{enumerate}[leftmargin=1.4em]
  \item  We present an {\it ordering-tag} framework that maps RC ordering rules to
    precise retirement-time and completion-time enforcement actions.
  \item We present a unified microarchitectural design in which the same
    ordering-tag abstraction governs ROB-head retirement predicates,
    merge-buffer completion, and selective load-hazard handling.
  \item We show that \method reduces geometric-mean normalized execution
    cycles by 7.9\% on native multithreaded workloads and improves
    geometric-mean IPC by 15.9\% on an instrumented SPEC2017 DBT proxy for
    cross-ISA execution (e.g., x86-on-Arm), while adding only 262 bytes of
    additional storage per core.
\end{enumerate}

The paper next reviews background, quantifies conventional ordering
overheads, presents \method and its correctness argument, compares against the
closest prior mechanism, and evaluates performance and hardware cost. We then
discuss TSO applicability and broader related work.


\section{Background}

\subsection{Definitions}
\label{sec:def}

\begin{figure}[h]
\includegraphics[width=0.45\textwidth]{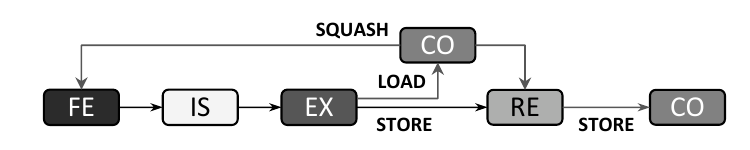}
\caption{Lifecycle of a memory access instruction. FE~(fetch); IS~(issue);
  EX~(execute); RE~(retire); CO~(complete).}
\label{fig:lifecycle}
\end{figure}

We use two terms throughout the paper. A memory instruction
\textit{completes} when its value or visibility event occurs: a load
completes when its value is returned, while a store completes when its
merge-buffer entry drains into the L1 data cache and becomes globally
visible. An instruction \textit{retires} when it leaves the reorder buffer
(ROB) head. Loads complete before retiring; stores can retire while their
committed store queue (SQ) entries await merge-buffer transfer and complete
when their merge-buffer entries drain. Thus, retirement can precede store
visibility.
Figure~\ref{fig:lifecycle} summarizes this timeline.

Program Order (PO) refers to the dynamic, sequential execution order of
instructions dictated by the program's control flow; \textit{po-older} and
\textit{po-younger} denote instructions that precede and follow a given
instruction in program order.

\subsection{Ordering Instruction Constraints}

\begin{figure}[h]
\centering
\includegraphics[width=\columnwidth]{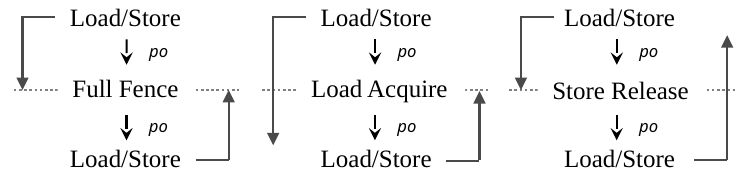}
\caption{Directional ordering constraints for full-fence, load-acquire, and
  store-release instructions.}
\label{fig:barriers}
\end{figure}

Ordering instructions enforce the constraints required by the memory model.
Under Release Consistency~\cite{gha:len90}, these instructions are either
\textit{bi-directional} (full fences such as \textit{mfence} or
\textit{dmb}) or \textit{uni-directional} (acquire/release operations).
A load-acquire constrains only po-younger operations, a store-release
constrains only po-older operations, and a full fence constrains both
directions. Armv8 exposes these semantics through \textit{ldar},
\textit{ldapr}, \textit{stlr}, and explicit fence instructions; the same
directional distinction is summarized in Figure~\ref{fig:barriers}.

We next summarize the microarchitectural structures relevant to ordering-instruction
execution.

\subsection{Merge Buffer}
\label{sec:merge_buffer}

A store can update the cache only when its target cache line is available.
After retirement, a committed store remains in the store queue until it can
transfer to a \textbf{merge buffer} (write buffer)~\cite{ska:cla97,ceb:jah24,
alv:ros19}. The path is \textit{store queue $\rightarrow$ merge buffer
$\rightarrow$ cache}. This decouples SQ deallocation from global visibility
and allows multiple stores to the same cache line to be merged into a single
merge-buffer entry. In our target design, the merge buffer is indexed by cache
line and holds at most one entry per cache line.

Only stores traverse the merge buffer; loads are serviced through the load
queue and cache hierarchy. Because store visibility is delayed until
merge-buffer drain, store-ordering instructions such as full fences and
store-release operations are directly coupled to drain behavior: if correctness
requires po-older stores to become globally visible first, retirement latency
becomes a function of merge buffer occupancy and cache-miss service time.

Unlike a TSO-style ordered store buffer, a release-consistency merge buffer
need not drain in FIFO order; entries may drain whenever their cache lines
are ready, except where explicit ordering instructions constrain
visibility. This is the flexibility that \method exploits. We next discuss
the resulting store-side and load-side ordering hazards.

\begin{figure}
\includegraphics[width=0.45\textwidth]{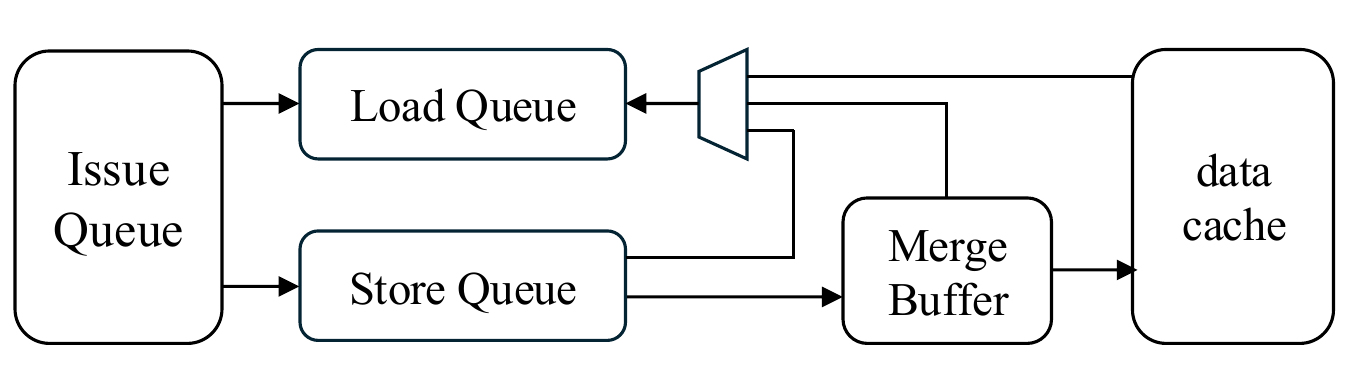}
\caption{Merge-buffer organization in the memory pipeline: loads and stores
  are issued through the load/store queues, while stores deallocated from
  the store queue enter the merge buffer before updating the data cache.}
\label{fig:merge_buffer}
\end{figure}

\subsection{Ordering Hazards}

Loads and stores are issued out of order~\cite{cai:lip04,gha:gup91}. A
po-younger load may receive data from a po-older same-address store in either
the store queue or the merge buffer (store-to-load forwarding, STLF), and
this can occur even across an intervening ordering instruction. Given this
behavior, we distinguish two hazard classes: store-side global-visibility
hazards and load-side invalidation hazards.

\subsubsection{Stores}

For stores, the hazard is a \textit{global-visibility} hazard: a po-younger
store must not become globally visible before a required po-older store. Under
release consistency, full fences order both sides of the fence, while a
store-release requires its po-older stores to become globally visible first.
Conventional implementations enforce these constraints by stalling retirement
until the required po-older merge-buffer entries drain.

\ignore{
Under release consistency, an unordered merge-buffer drain is legal by
default and must be constrained only where explicit synchronization requires
ordering. Under TSO, FIFO drain preserves baseline store ordering and thus
reduces the scope of additional fence-induced constraints.
}

\subsubsection{Loads}

For loads, the hazard arises when a speculative load completes while a
po-older full fence or load-acquire is still active, and a later invalidation
shows that the loaded value may no longer satisfy that ordering constraint.
Conventional implementations track such loads in the load queue, mark them
hazardous on a matching snoop, and defer squash/replay until the po-older
ordering instruction retires. This hazard-tracking path is required for full
fences and load-acquires, but not for store-release, which does not order
po-younger loads.

\ignore{
The same hazard-tracking path also handles a distinct recovery case: two
loads to the same address may complete out of age order, with the younger
load completing first. Because accesses to the same address must remain
sequentially consistent, completion of the older load triggers
squash/replay of the younger load if that younger load was marked hazardous.
In this case, the younger load may have read a stale value, whereas the
older load observes the newer value.
}

\begin{table}[t]
  \centering
  \footnotesize
  \setlength{\tabcolsep}{3pt}
  \caption{Retirement-time constraints in a conventional
    Release-Consistency implementation.}
  \begin{tabular}{@{}p{0.32\columnwidth}p{0.24\columnwidth}p{0.30\columnwidth}@{}}
    \toprule
    Ordering instruction & Merge-buffer drain & Hazardous-load squash \\
    \midrule
    load-acquire & Not required & Required \\
    store-release & Required & Not required \\
    full fence & Required & Required \\
    \bottomrule
  \end{tabular}
  \label{tab:overhead}
\end{table}

Table~\ref{tab:overhead} summarizes the fence-retirement actions in a
conventional design. These constraints directly motivate the inefficiencies
analyzed in Section~\ref{sec:inefficiency}: speculation-window shrinkage,
drain-induced store-retirement stalls, and unnecessary load squash/flush.
We next state the Release-Consistency semantics used in the remainder of the
paper; Section~\ref{sec:versioning} then maps these semantics to our
tag-based framework.

\subsection{Release-Consistency Semantics}
\label{sec:rc_rules}

The following semantic rules describe the subset of Armv8
\textbf{Release Consistency} that TEMPO must enforce for the mechanisms
studied in this paper~\cite{flu:gra16,pul:flu17}. We intentionally use a
simplified operational model rather than the full Armv8 consistency
model. This abstraction captures the acquire/release/full-fence ordering
constraints and load-value constraints exercised by our microarchitecture;
features outside that scope are not modeled here.

Here, $X$ and $Y$ denote two instructions; \mpo and \mvo denote program order
and global visibility order, respectively. $L(a)$ and $S(a)$ denote a load
from and a store to address $a$, respectively. $FF(X)$ is true if $X$ is a full fence,
$LDAR(X)$ is true if $X$ is an $RC_{SC}$ acquire load, where $RC_{SC}$
denotes release consistency with sequentially consistent acquire semantics;
$LDAPR(X)$ is true if $X$ is an $RC_{PC}$ acquire load, where $RC_{PC}$
denotes release consistency with processor-consistent acquire semantics;
$STLR(X)$ is true if $X$ is a
store-release instruction.

For any instruction pair $(X, Y)$ such that $X \po Y$, the following rules
apply:

\begin{enumerate}[label=RC\arabic*.,leftmargin=1.25cm]
\item $ FF(X)\;or\;FF(Y) \implies X \vo Y $
\item $ LDAR(X)\;or\;LDAPR(X) \implies X \vo Y $
\item $ STLR(Y) \implies X \vo Y $
\item $ X, Y \in (FF,\; LDAR,\; STLR) \implies X \vo Y $
\item $ X, Y \in (L(a),\; S(a)) \implies X \vo Y $
\item $ Value\; of\; L(a) = Value\; of\; Max_{<v}\; \{S(a)\; |\\
  S(a) \po L(a)\; or\; S(a) \vo L(a)\} $
\end{enumerate}

In $RC$, unlike in sequential consistency ($SC$), program order does not
imply global order for ordinary memory accesses. A synchronizing load
(\textit{load-acquire}), a synchronizing store (\textit{store-release}), and a
fence instruction enforce order among the memory accesses.

$RC1$ states that a full fence orders all po-older accesses before all
po-younger accesses.

$RC2$ states the forward-ordering property of acquire
loads: both $LDAR$ and $LDAPR$ are ordered before po-younger memory accesses.

$RC3$ states that a store-release instruction is ordered after all po-older memory accesses.

$RC4$ places $FF$, $LDAR$, and $STLR$ in a single sequential-order class.
Thus, although both $LDAR$ and $LDAPR$ impose forward ordering on po-younger
operations through $RC2$, only $LDAR$ participates in the stronger $RC4$
ordering relation with po-older $STLR$ instructions. In other words, $LDAPR$
need not be sequentially consistent with a po-older store-release instruction
and can be freely reordered across such an instruction.

\ignore{
$RC4$ captures the stronger sequential-ordering requirement of $LDAR$:
unlike $LDAPR$, it is sequentially consistent with $STLR$ and full-fence
instructions.
}

Cross-core behavior is captured through coherence and read-from relations,
reflected in $RC5$ and $RC6$. $RC5$ requires accesses to the same address
to preserve program order in global visibility order. In this abstraction,
$RC5$ captures the per-location sequential-consistency requirement for
same-address accesses.

$RC6$ is the load-value (read-from) rule: the value of a load is
obtained from the latest po-older or globally ordered store to the same
address. Building on this rule, a load first checks the store queue and then the
merge buffer for a po-older overlapping store; if no such store exists, it
reads from the cache hierarchy, which provides the latest globally visible
value under the Armv8 multi-copy-atomic model~\cite{pul:flu17}.

\section{Motivation}
\label{sec:inefficiency}

\begin{figure*}
\begin{centering}
\includegraphics[width=0.9\textwidth]{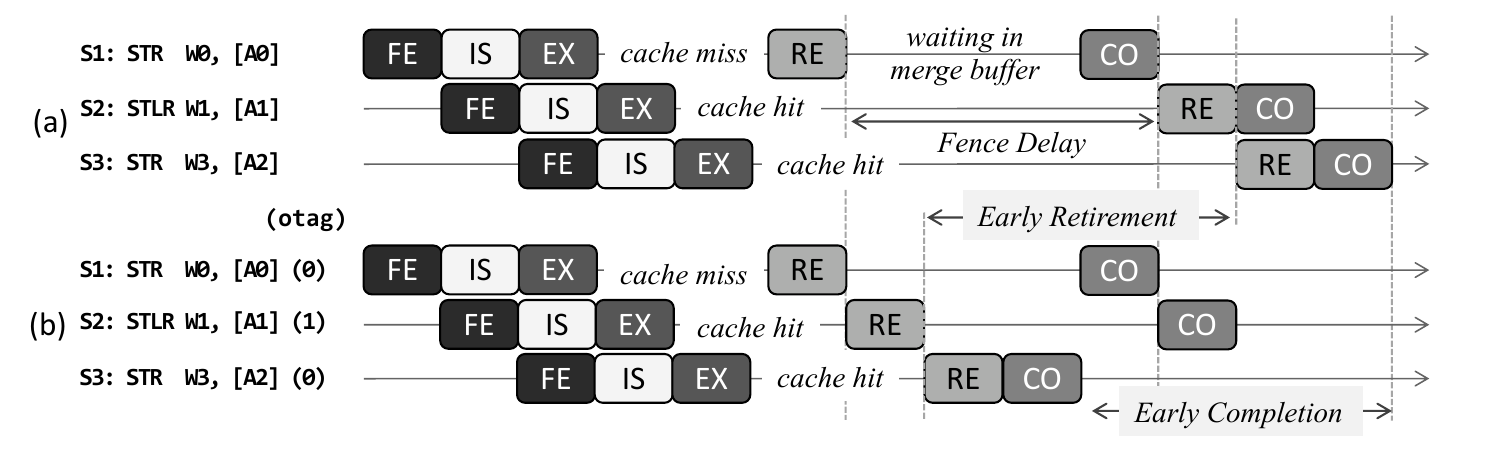}
\caption{Early retirement and completion of stores from an unordered merge
  buffer: (a) baseline and (b) with \method. In (b), the store-release $S2$
  receives its own release tag, while ordinary stores $S1$ and $S3$ retain
  the current global tag. FE~(fetch); IS~(issue); EX~(execute); RE~(retire);
  CO~(complete); \textit{otag}~(ordering tag).}
\label{fig:early_store_retirement}
\end{centering}
\end{figure*}

The retirement-time constraints summarized in
Table~\ref{tab:overhead} manifest as three concrete inefficiencies in a
conventional design.

In conventional implementations, ordering instructions introduce three
costs: (i) retirement backpressure that shrinks the effective speculation window, 
(ii) drain-induced retirement stalls while required po-older
stores wait in the merge buffer, and (iii) unnecessary load-side
squash/flush after matching invalidations.
Together, these overheads reduce performance and waste energy: stalled
cycles consume dynamic power without making forward
progress, and unnecessary squashing and replaying waste work that was already
correctly executed.
We analyze these inefficiencies in the rest of this section.

\subsection{Shrinking Speculation Window}

The effective speculation window is bounded by available ROB, load-queue,
and store-queue capacity. In a conventional design, when a fence reaches the
ROB head, retirement can stall while waiting for required merge-buffer drain.
During this interval, younger instructions cannot retire, and load/store
queue entries associated with completed memory operations cannot be
reclaimed. Because speculative loads and stores also consume these entries,
the load/store queue can eventually become saturated.

This backpressure reduces the number of younger memory operations that can be
issued, shrinking the effective speculation window. As queue pressure grows,
the core loses memory-level parallelism and increasingly exposes
fence-induced latency. We quantify this effect later through the load/store
queue full-stall results in Figure~\ref{fig:native_lqsq_full}.

This speculation-window shrinkage is the aggregate effect observed by the
pipeline. We next isolate the two mechanism-level causes in conventional
ordering-instruction handling: drain-induced retirement stalls and unnecessary
load squash/replay after invalidation matches.

\subsection{Inefficient Store Retirement} 
\label{sec:store_ret}

A full fence or store-release at the ROB head must wait until required
po-older stores drain from the merge buffer to preserve visibility
order. Because drain latency is determined by pending cache-miss service for those
stores, this wait serializes retirement, increases latency for following
instructions, and can propagate backpressure to instruction fetch when the
ROB fills up.

A store-release does not constrain po-younger memory accesses. In
Figure~\ref{fig:early_store_retirement}, stores $S1$, $S2$, and $S3$ are
independent, and $S2$ is a store-release. In
Figure~\ref{fig:early_store_retirement}(a), $S2$ cannot retire until the
merge-buffer entry of po-older store $S1$ drains. Because retirement is
in order, this also prevents $S3$ from passing $S2$ in the ROB, even though
$S2$ imposes no ordering constraint on $S3$. If $S1$ is a cache miss, the
resulting delay is long. Once $S1$ completes, both $S2$ and $S3$ retire, but
$S3$ could have completed earlier without violating the memory model.

With \method, stores $S1$, $S2$, and $S3$ retire at the ROB head and later
transfer from the store queue to the merge buffer. Their ordering tags ensure that
$S2$ completes only after $S1$, while imposing no additional constraint on
$S3$. In our design, tag assignment happens at decode, which is in program
order. Because a store-release orders only po-older instructions, $S2$
receives a new release tag, but the po-younger ordinary store $S3$ keeps the
current global tag, as described in
Section~\ref{subsec:ordering-tags} and illustrated in
Figure~\ref{fig:ordering_tags}. $S3$ can therefore complete before $S2$ even
though $S2$ is the intervening store-release. As shown in
Figure~\ref{fig:early_store_retirement}(b), $S3$ therefore completes earlier.
Retiring $S2$ and $S3$ without waiting
also removes a critical path delay for following instructions.

For full fences, \method preserves the required constraint that po-younger
stores do not complete before po-older stores, while still decoupling fence
retirement from merge-buffer drain. Section~\ref{sec:versioning} explains how
this retirement/completion decomposition avoids the corresponding bottlenecks.

\begin{figure}[t]
\begin{centering}
\includegraphics[width=0.75\columnwidth]{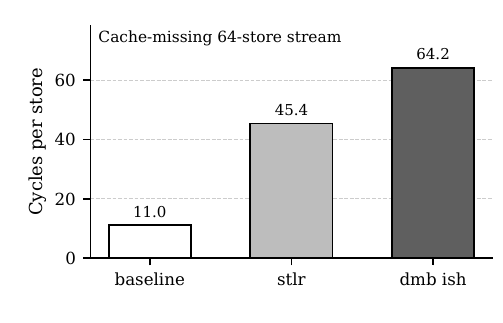}
\caption{Store-stream microbenchmark showing the overhead of store-side
ordering instructions. Bars show a cache-missing single-thread stream of
64 stores, where each operation is either a regular store plus \texttt{nop},
\texttt{stlr} plus \texttt{nop}, or a regular store plus \texttt{dmb}.}
\label{fig:store_stream_native}
\end{centering}
\vspace{-1em}
\end{figure}

\begin{figure}[t]
\begin{centering}
\includegraphics[width=0.84\columnwidth]{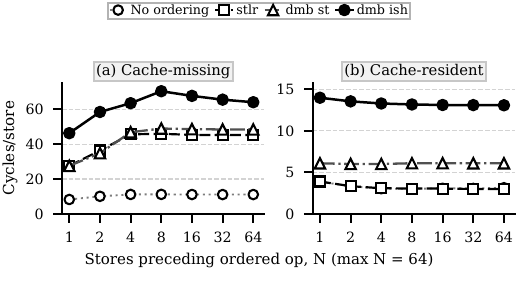}
\caption{Native store-stream latency with cache-missing and cache-resident
stores. The experiment sweeps the number of stores preceding each ordering
operation and reports cycles per store.}
\label{fig:store_stream_miss_hit_native}
\end{centering}
\vspace{-1em}
\end{figure}

Figure~\ref{fig:store_stream_native} isolates the ordering-instruction
overhead on Neoverse-V2 hardware~\cite{nvidia-gh200}. In this cache-missing
64-store stream, the regular-store+\texttt{nop} baseline sustains about
$11$ cycles/store, whereas \texttt{stlr}+\texttt{nop} costs $45.4$
cycles/store and regular-store+\texttt{dmb ish} costs $64.2$
cycles/store. This directly shows how a conventional implementation can
serialize a cache-missing store stream around store-side ordering
instructions when older stores are still pending.
Figure~\ref{fig:store_stream_miss_hit_native} separates the same effect for
cache-missing and cache-resident streams.  The cache-missing case shows
that ordering latency grows with outstanding older stores, while the
cache-resident case isolates the fixed cost of each ordering instruction.

\subsection{Inefficient Load Speculation}
\label{sec:load_spec}

\begin{figure*}
\begin{centering}
\includegraphics[width=0.9\textwidth]{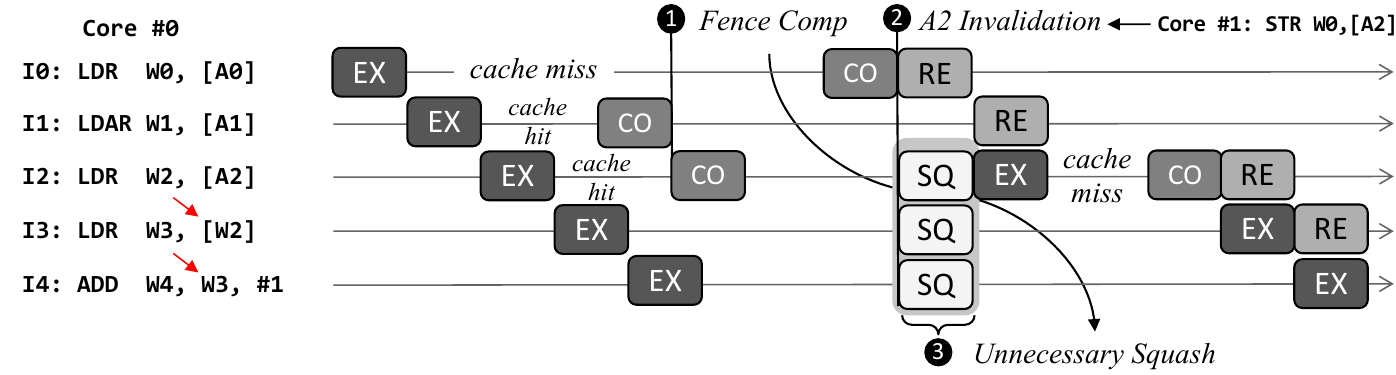}
\caption{Inefficient load speculation in a weak consistency architecture.
  EX~(execute); RE~(retire); CO~(complete); SQ~(squash) in this timeline.}
\label{fig:load_speculation}
\end{centering}
\end{figure*}

In the conventional implementation considered here, incoming coherence
invalidations are matched against speculatively completed loads in the load
queue, and matching loads are marked hazardous.

The inefficiency is unnecessary squash-and-replay after such matches. Under
Release Consistency, recovery is needed only while a relevant po-older
ordering constraint remains active for the load: for a load-acquire, until the
acquire completes; for a full fence, until the required lower-tag stores have
drained. Once that constraint is no longer active, the same invalidation no
longer implies an ordering violation and the load need not even remain
hazardous. We illustrate the load-acquire case next.

In Figure~\ref{fig:load_speculation}, instruction $I1$ is a load-acquire
access to address $A1$. Instructions $I0$ and $I2$ are independent loads to
$A0$ and $A2$, respectively, and $I3$ and $I4$ depend on the value loaded
by $I2$. The example assumes that $I0$ is a cache miss, whereas $I1$ and
$I2$ are cache hits. Consequently, $I1$ and $I2$ complete speculatively
before $I0$ at \mynode{1}, while $I0$ is still waiting for its cache
line. At \mynode{2}, an invalidation for the cache line containing $A2$
arrives and marks $I2$ hazardous in the load queue. In a conventional
implementation, $I2$ is squashed, along with its dependents $I3$ and $I4$,
when $I1$ retires. However, because the load-acquire $I1$ has already
completed at \mynode{1}, the ordering it enforces on $I2$ is satisfied at
this point. Therefore, squashing and replaying $I2$, $I3$, and $I4$ is
unnecessary.

These inefficiencies in conventional implementations motivate the
\method mechanism presented in Section~\ref{sec:versioning}, which
decouples retirement-time stalling from completion-time ordering and
enables selective load-hazard handling. The resulting end-to-end performance
impact appears later in Figure~\ref{fig:native_exec_cycles}.


\section{TEMPO Design}
\label{sec:versioning}

We first define ordering tags and their decode-time assignment, then describe
the required microarchitectural augmentations. The remaining subsections
present retirement predicates, tag-ordered store completion, selective
load-hazard handling, and finally correctness intuition.

\subsection{Ordering Tags}
\label{subsec:ordering-tags}

\method assigns an ordering tag to each memory instruction at decode so that
ordering enforcement can be split across retirement and completion. In our
design, decode is in program order, so assigning tags there is sufficient to
preserve program-order semantics. More generally, any implementation with
out-of-order decode can assign tags at the point where it assigns sequence
numbers or equivalent program-order identifiers to instructions. Tags let
the core enforce store-side constraints by comparing tags at merge-buffer
completion, while load-side constraints are enforced through tag-aware
retirement predicates. This decoupling avoids retirement-time drain stalls
without weakening required ordering.
Operationally, tags convert global ordering into a local monotonic predicate:
lower-tag stores must become visible before higher-tag stores can release
unsafe loads that depend on them.

Tag assignment follows ordering-instruction semantics. At decode, each
ordinary load or store is assigned the current value of the
\textit{\textbf{global tag counter}} $g$ as its ordering tag. The decode unit
also maintains a store-release tag register $r$, which holds the most
recently assigned store-release tag. When decoding a store-release, the
decode unit updates $r \leftarrow \max(g,r)+1$ and assigns that new value to
the store-release instruction's ordering tag. The global counter $g$ is
unchanged, so po-younger ordinary accesses continue to inherit the current
value of $g$; this isolates the synchronizing store without over-constraining
intervening regular operations. The update rule ensures that consecutive
store-release instructions receive strictly increasing tags even when $g$ is
unchanged. When decoding a full fence, the decode unit updates
$g \leftarrow \max(g,r)+1$; the fence itself has no ordering tag, but
subsequent ordinary accesses inherit this larger value of $g$. This asymmetry is the key
decode-time insight: store-release must be ordered after po-older stores, but
it need not impose a new tag boundary on po-younger ordinary accesses. By
contrast, a full fence does advance $g$, creating a hard tag boundary on both
sides of the fence.

A load-acquire leaves both $g$ and $r$ unchanged at decode; its constraints are
enforced through retirement-time completion gating rather than at decode. A
load-acquire's completion signal is held while a required lower-tag store
remains in the store queue or merge buffer; retirement then checks younger
invalidation-marked hazardous loads
(Table~\ref{tab:enforcement_timing}, Section~\ref{subsec:retirement}). To
support $RC4$ under $RC_{SC}$ (the stronger \textit{LDAR}/$RC_{SC}$ acquire, as
opposed to the weaker \textit{LDAPR}/$RC_{PC}$ acquire), a load-acquire's completion signal is also
held while any po-older store-release remains in either structure; an
$RC_{PC}$ load-acquire does not impose that additional completion delay.

Ordering tags encode completion-time store-side ordering; they do not encode
every load-side ordering edge. In particular, load-acquire ordering and the
$RC4$ constraint under $RC_{SC}$ are enforced by retirement-time checks
rather than by tag monotonicity alone. Figure~\ref{fig:ordering_tags}
summarizes the decode-time updates.

\begin{figure}[t]
\centering
\footnotesize
\setlength{\tabcolsep}{3pt}
\renewcommand{\arraystretch}{1.05}
\newcommand{\taginst}[2]{\texttt{\makebox[3.8em][l]{#1}#2}}
\newcommand{\taginstbf}[2]{\textbf{\texttt{\makebox[3.8em][l]{#1}#2}}}
\begin{tabular}{@{}p{0.24\columnwidth}p{0.10\columnwidth}p{0.20\columnwidth}p{0.34\columnwidth}@{}}
\toprule
\textbf{Instruction} & \textbf{otag} & \textbf{Type} & \textbf{Update} \\
\midrule
\multicolumn{4}{@{}l@{}}{\textit{Initial state:} $g=0,\; r=0$} \\
\addlinespace[0.25em]
\taginst{STR}{[A1]} & 0 & Ordinary & inherit $g=0$ \\
\taginstbf{STLR}{[A2]} & \textbf{1} & \textbf{Store-release} & $r \leftarrow \max(g,r)+1$ \\
\taginst{STR}{[A3]} & 0 & Ordinary & inherit $g=0$ \\
\taginstbf{STLR}{[A4]} & \textbf{2} & \textbf{Store-release} & $r \leftarrow \max(g,r)+1$ \\
\addlinespace[0.35em]
\taginstbf{DMB}{ISH} & -- & \textbf{Full fence} & $g \leftarrow \max(g,r)+1$ \\
\taginst{STR}{[A5]} & \textbf{3} & Ordinary & inherit $g=3$ \\
\addlinespace[0.35em]
\taginstbf{LDAR}{[A6]} & 3 & \textbf{Load-acquire} & no tag update \\
\taginst{LDR}{[A7]} & 3 & Ordinary & inherit $g=3$ \\
\bottomrule
\end{tabular}
\caption{Ordering tag update across a continuous instruction stream. Ordinary
accesses inherit the current global tag counter $g$. Store-release
instructions advance $r$ and receive strictly increasing release tags; a full
fence advances $g$ to exceed any prior store-release tag, and a load-acquire
leaves tags unchanged at decode.}
\label{fig:ordering_tags}
\end{figure}

Two implementation details matter for realizing this scheme in hardware:
correctly recovering the decode-side tag state across flushes and sufficient
tag width to avoid wraparound comparison ambiguity.

Because ordering tags are assigned before retirement, a pipeline flush must
restore the two decode-side tag registers, $g$ and $r$, to the values they
held at the recovery point. In practice, the processor snapshots these two
registers alongside the other rename/sequence state used for flush recovery
and restores them when the flushed younger instructions are discarded. This
ensures that redecoded instructions receive the same program-order tag state
they would have seen without the mis-speculation.

The ordering-tag width must cover all in-flight memory instructions in the
instruction window, including pipeline-resident instructions. For a
$2^{N}$-entry in-flight window, we use an $N+1$-bit ordering tag. This
prevents wrap ambiguity while those instructions remain in flight and enables
\textit{wraparound logic} to compare tags correctly when the counter wraps
\cite{rfc1982, cum02}.

As a concrete example, consider $N{=}3$: the in-flight bound is 8, so tags are
4-bit values (0--15). If active tags are $\{14,15,0,1\}$, wraparound logic
computes modular distance $\Delta(a,b)=(a-b)\bmod 16$ and treats $a$ as
younger than $b$ when $0<\Delta(a,b)<8$. Thus, tag 0 is correctly ordered
after tag 15, and tag 1 after tag 0, while ambiguity with much older
generations is excluded by the 8-entry in-flight bound.

At a high level, TEMPO uses tags to move store-side ordering to completion
time while keeping load-side ordering at retirement time. We next extend this
mechanism to directional barriers before describing the microarchitecture.


\subsection{Supporting Load-Only and Store-Only Barriers}
\label{subsec:directional-barriers}

The base mechanism above targets full barriers together with acquire and
release operations. If the ISA also exposes directional barriers such as
\texttt{dmb ld} and \texttt{dmb st}, we can extend the decode-time assignment
logic while still keeping exactly one \textit{ordering tag} per load and one
\textit{ordering tag} per store.

The decode unit then maintains three registers: a \textit{global store tag
counter} $g$ for ordinary stores, a \textit{load assignment register} $l$ for
ordinary loads, and the store-release tag register $r$. Ordinary stores
inherit $g$ as their ordering tag, while ordinary loads inherit $l$ as their
ordering tag. A full fence computes $t \leftarrow \max(g,r)+1$ and updates
$g \leftarrow t$ and $l \leftarrow t$, so po-younger stores receive a new store
ordering region and po-younger loads wait for all lower-tag stores before
retiring. A store-release still receives $\max(g,r)+1$ and updates only $r$.

A store-only barrier advances only $g$. This separates later stores from
earlier stores without forcing later loads to wait for those earlier stores.
Conversely, a load-only barrier leaves both $g$ and $l$ unchanged. Younger
loads already retire after older loads complete, so ld-ld ordering does not
require a new load tag boundary. When the load-only barrier retires, the
core only needs to check for younger hazardous loads and delay retirement
until any such load-side hazard is resolved.

Under this extension, a load's ordering tag acts as a store-drain threshold:
a load with ordering tag $t$ waits only for pending po-older stores whose
ordering tag is \emph{less than} $t$. Stores with the same ordering tag do not
by themselves delay that load. This preserves the intended directional
semantics: \texttt{dmb st} orders only st-st, \texttt{dmb ld} orders only
ld-ld, and a full \texttt{dmb} orders both.

\ignore{

As a concrete \texttt{dmb st} example, consider
\texttt{st [x]; dmb st; ld [y]; st [z]}. If the pre-barrier state is
$g{=}l{=}0$, the first store receives ordering tag 0, \texttt{dmb st}
advances only $g$ to 1, the load still receives ordering tag 0 from $l$, and
the final store receives ordering tag 1. Consequently, the final store cannot
drain before the first store, so the required st-st ordering is preserved.
However, the intermediate load need not wait for the first store, because its
ordering tag is 0 and loads wait only for po-older stores with strictly
smaller ordering tags. This avoids conservatively stalling a load that the
store-only barrier does not need to order.

Similarly, for \texttt{ld [x]; st [y]; dmb ld; ld [z]}, the younger load is
already prevented from retiring before the older load completes by in-order
load retirement. Therefore, \texttt{dmb ld} need not create a new tag
boundary or force the younger load to wait for the intervening store;
retirement only checks for younger hazardous loads before allowing the
barrier to retire.
}

\subsection{Microarchitecture}
\label{subsec:microarchitecture}

\method augments the load queue, store queue, and merge buffer with an
ordering-tag field per entry. Store-side ordering is enforced at merge-buffer
drain time: an entry may drain only when its cache line is available and its
ordering tag is the lowest among in-flight merge-buffer tags.

In a conventional design, the merge buffer is indexed by cache line and
therefore holds at most one entry per cache line. In \method, the
merge-buffer key is extended to \textit{(ordering tag, cache-line address)}.
As a result, same-line stores with different ordering tags can coexist as
distinct entries, while same-line stores with the same ordering tag can still
merge. This allows younger same-address stores to continue transferring into the
merge buffer without waiting for an older differently tagged same-line entry
to drain. It does not forfeit a cross-boundary coalescing opportunity that the
baseline would have realized: when two same-line stores are separated by an
ordering boundary, the younger store remains in the store queue until the
older ordered store drains, so the two stores are not simultaneously
merge-buffer-resident there either.

As in conventional coalescing designs, each merge-buffer entry may observe a
short merge-wait interval before becoming drain-eligible, allowing additional
same-tag, same-line stores from the store queue to merge. At each completion
opportunity, an entry is eligible only if this wait has expired, the cache
line is ready, and the entry belongs to the current minimum in-flight
ordering tag (with the oldest-entry override in
Section~\ref{subsec:completion}).

To track the minimum in-flight merge-buffer ordering tag efficiently, we
maintain an \textit{ordering-tag queue} that tracks tags with resident
merge-buffer entries. Each entry stores the current number of merge-buffer
entries for that tag and a one-bit fence bit. When a new merge-buffer entry is
allocated, the corresponding count for the tag is incremented; on merge
into an existing entry, it is unchanged; and on drain, it is
decremented. The queue stores distinct active tags in modularly sorted order
and maintains the current minimum tag at its head. Allocations increment a
matching entry or insert a new tag at its sorted position. Drains decrement the
matching tag, normally the head but potentially a non-head tag under the
oldest-entry override. Non-head drains leave the queue's sorted order unchanged.
A full fence
sets the bit on its maximum pre-fence tag, $\max(g,r)$ (tag 2 in
Figure~\ref{fig:ordering_tags}). A fence-marked tag is ineligible for the
oldest-entry override and drains only at the head. Thus, when its count reaches
zero, all lower tags have drained, so the core checks the load queue for
hazardous loads before advancing. The next
non-empty tag then becomes drain-eligible. This enforces global visibility
in tag order while retaining the flexibility of an unordered merge buffer.
Because the ordering-tag queue tracks only non-speculative stores that
already reside in the merge buffer, it does not require flush-time repair:
only retired stores transfer from the store queue to the merge buffer, so
speculative stores never enter either the merge buffer or the ordering-tag
queue. Any store represented in the ordering-tag queue is therefore already
retired and cannot later be flushed. When an unsafe load reaches the ROB head,
its completion predicate checks both structures: no po-older lower-tag store
may remain in the store queue, and the load's tag must be less than or equal to
the current minimum merge-buffer tag. These checks reuse the load-store unit's
existing completion signal and require no new merge-buffer-to-ROB channel.

\begin{table*}[t]
\centering
\small
\caption{Timing of ordering-enforcement actions in baseline and \method designs.}
\label{tab:enforcement_timing}
\begin{tabular}{p{0.23\textwidth}p{0.35\textwidth}p{0.35\textwidth}}
\toprule
Ordering action & Baseline (when enforced) & \method (when enforced) \\
\midrule
Full fence: store-side ordering &
At full-fence retirement, required po-older stores drain from the merge buffer before the fence retires. &
The full fence retires at the ROB head; required store ordering is enforced later by tag-based order at merge-buffer drain. \\
\midrule
Full fence: load-hazard handling &
After required store drain at fence retirement, the load queue is checked and hazardous loads are squashed/replayed. &
Deferred until the store queue has no relevant lower-tag store and the fence-associated ordering-tag count reaches zero; the load queue is then checked and hazardous loads are squashed/replayed. \\
\midrule
Invalidation: hazard-marking condition &
Matching invalidations can mark speculative loads hazardous conservatively. &
A matching invalidation marks a speculative load hazardous if either (i) a
po-older load-acquire is active (incomplete), or (ii) po-older lower-tag
stores are still pending in the store queue or the merge buffer. \\
\midrule
Store-release: store-side ordering &
Store-release retirement waits until required po-older stores drain from the merge buffer. &
Store-release retires at the ROB head; required ordering with po-older stores is enforced by tag-based order at merge-buffer drain. \\
\bottomrule
\end{tabular}
\end{table*}

Table~\ref{tab:enforcement_timing} summarizes, for each ordering-related
action, when enforcement occurs in the baseline and in \method. We next
define the detailed retirement and completion rules.

\subsection{Retirement}
\label{subsec:retirement}

This subsection defines the architectural retirement conditions under
\method.

We use the \textit{safe}/\textit{unsafe} terminology of prior
work~\cite{sin:nar13}, but in \method the classification is implemented
directly in hardware from the realized value source, specifically the
ordering tag of the forwarding store when STLF occurs, without separate
static or dynamic analysis. This execution-time predicate directly defines
ROB-head retirement: safe accesses retire without waiting for merge-buffer
drain, whereas unsafe accesses retire only after their required ordering
conditions are satisfied.

\noindent\textbf{Stores.}
All stores are \textit{safe}. At the ROB head, a store retires by becoming
non-speculative in the store queue and may remain there until merge-buffer
transfer succeeds; backpressure delays SQ deallocation, not retirement. After
transfer, visibility is governed by tag-based merge-buffer completion
(Section~\ref{subsec:completion}).

\noindent\textbf{Loads.} Loads are classified as \textit{safe} or
\textit{unsafe}. A load is \textit{safe} iff it receives store-to-load
forwarding (STLF) from the youngest matching po-older store with the same
ordering tag, after probing the store queue and then the merge buffer. Thus,
unlike prior schemes that classify loads through separate static or dynamic
analyses, \method classifies a load directly from its realized value source.
A safe load may retire at the ROB head even if the forwarding store has not
yet drained from the merge buffer. This reflects local visibility: a load may
consume data from a po-older local store before that store becomes globally
visible, unless an intervening ordering instruction enforces additional
constraints~\cite{sew:sar10,alg:mar14,nag:sor11}.

An \textit{unsafe} load is any load that either receives STLF from a
lower-tag po-older store or obtains its value from the cache hierarchy. At the
ROB head, all po-older stores have retired, but some may still await
merge-buffer transfer in the store queue. The load may retire only when no
required po-older lower-tag store remains in either structure: no committed
lower-tag SQ entry exists, and the load's tag does not exceed the minimum
merge-buffer tag. \method reuses the load-store unit's completion signal,
holding it until both checks succeed without adding a ROB-stage comparator.
This delay is necessary because, unlike a \textit{safe}
load, an \textit{unsafe} load does not establish that the required po-older
store ordering has already been respected. If such a load retired while a
po-older lower-tag store were still pending in either structure, the machine
could commit the younger load before the older ordered store had become
visible, violating the full-fence ordering rule ($RC1$) and potentially the
load-value rule ($RC6$). For example, in \textit{store A; full fence; load B},
\textit{load B} cannot retire while lower-tag \textit{store A} remains in
either structure.

This classification is distinct from \textit{hazardous} loads in
Section~\ref{sec:spec_exec_inv}. \textit{Safe/unsafe} determines whether a
load may retire with respect to po-older store visibility, whereas
\textit{hazardous} denotes a speculative load that matched an invalidation and
may require deferred squash/replay at an ordering trigger.

\noindent\textbf{Ordering instructions.}
A full fence or store-release instruction may retire at the ROB head without
waiting for the merge buffer to drain; the required store-side ordering is
enforced later, at merge-buffer completion. This differs from a TSO-style
ordered merge buffer, where a fence need not block retirement waiting for
drain because store visibility is already ordered by the buffer itself. Under
release consistency, however, the merge buffer is unordered, so a
conventional design must enforce store ordering explicitly by waiting for the
pending merge-buffer stores to drain. \method{}~preserves that ordering without
that retirement-time drain by enforcing it through tag-ordered drain.
A load-acquire follows the same
\textit{safe}/\textit{unsafe} retirement predicate as any other load and must
also complete the retirement-time check for younger invalidation-marked
hazardous loads, squashing/replaying them if needed. Under $RC_{SC}$, however, an
additional completion constraint applies ($RC4$): the load-acquire's
completion signal is held while any po-older store-release remains in either
structure. This delayed completion applies to $RC_{SC}$ acquires such as
\textit{LDAR}. By contrast, an $RC_{PC}$ acquire such as \textit{LDAPR} does
not impose that additional completion delay and may retire once the general
load predicate and this retirement-time hazardous-load check are satisfied.

\noindent\textbf{Atomics (CAS/AMO).}
\method does not alter the baseline atomicity mechanism of read-modify-write
operations; it only maps their ordering attributes to the same enforcement
rules. Relaxed atomics use the current ordering tag. Acquire atomics follow
load-acquire retirement and retirement-time hazardous-load replay rules,
release atomics follow
store-release tag/completion rules, and acq\_rel (or stronger) atomics apply
both.

Having defined when instructions may retire, we next define how stores become
globally visible under ordering-tag control.

\subsection{Store Completion}
\label{subsec:completion}

When a retired store transfers from the store queue into the merge buffer,
the core issues a prefetch request
for exclusive ownership of the target cache line. The merge-buffer entry becomes
completion-eligible after its merge-wait interval expires and the prefetched
line is ready in L1. Because the merge buffer is physically unordered,
completion is selected by ordering tag rather than physical position: among
eligible entries, the core selects from the current minimum in-flight tag group
(subject to the oldest-entry override below). This preserves required
tag-based global-visibility order while retaining out-of-order completion
across independent cache lines.

To ensure liveness, especially for high-tag store-release entries, we add an
\textit{oldest-entry override}: when the oldest merge-buffer entry is
cache-ready, it may complete regardless of ordering tag unless its tag carries
a fence bit. This is safe because
no po-older merge-buffer entry remains, so completion cannot violate required
store ordering under the full-fence and store-release rules ($RC1$ and
$RC3$). Because retired stores transfer from the store queue in age order,
stores still resident there are younger than all current merge-buffer entries
and cannot become globally visible until they enter the merge buffer.

Merge-buffer entries are allocated as retired stores transfer from the
store queue in age order,
so all valid entries already present when a new entry is allocated are older
than that new entry. The oldest-entry condition can therefore be implemented by
a snapshot bit-vector of all valid entries captured at allocation. Each
draining entry clears its bit; when the vector reaches zero, the entry is the
oldest and may drain without tag restriction. In effect, the vector indicates
when the entry becomes the oldest.

As retired stores leave the store queue and merge-buffer entries complete, the
two lower-tag predicates clear and unblock waiting unsafe loads. The next
subsection completes the enforcement flow by describing deferred load-hazard
handling.

\subsection{Load Hazards}
\label{sec:spec_exec_inv}

An invalidation that matches a speculative load indicates an ordering
hazard only while the load is still constrained by a po-older ordering
instruction.  For full-fence constraints, \method checks whether a po-older
lower-tag store remains pending in either the store queue or the merge
buffer; for load-acquire constraints, a po-older load-acquire is considered
active only until that acquire completes.  In a conventional baseline, this
relation is not explicit, so matching loads are conservatively marked
hazardous. In \method, hazard marking is avoided when neither condition
holds. \textit{Safe} loads are also exempt: their value came from same-tag
forwarding from a po-older local store, so the matched invalidation does
not indicate a violation of the relevant ordering rule ($RC1$ for
full-fence constraints, $RC2$ for load-acquire constraints).

For each matching invalidation, the core checks two conditions: whether
lower-tag stores remain pending in the store queue or the merge buffer, and
whether a po-older load-acquire remains incomplete. If either holds, the
load is marked hazardous. At the corresponding trigger point, the core
checks the load queue and squashes/replays such loads if needed: at
load-acquire retirement, or, for full-fence constraints, when no relevant
lower-tag store remains in the store queue and the corresponding fence-bit
count reaches zero in the ordering-tag queue. This
deferred replay prevents speculative loads from becoming architecturally
visible in violation of $RC2$ (for active load-acquire constraints) or
$RC1$ (for pending lower-tag full-fence constraints). Read-after-read (RAR)
hazards from same-address load reordering are handled identically in the
baseline and in \method: they reuse the same invalidation-marked
hazardous-load path, but are triggered when an older load to the same
address completes after a younger load. The proposed mechanism changes only
the invalidation-based handling for active full-fence and load-acquire
ordering constraints.

With retirement, completion, and deferred hazard handling defined, we close by
summarizing why these enforcement points suffice for correctness.

\subsection{Correctness Intuition}
\label{subsec:correctness-intuition}

\method splits enforcement across ROB retirement and merge-buffer
completion. Retirement predicates (\textit{safe}/\textit{unsafe} load
classification, load-acquire activity tracking, and deferred hazard
processing) govern when instructions may retire, while completion enforces
tag-ordered store visibility at merge-buffer drain. Correctness follows from
the composition of these enforcement points rather than from either one alone.

The key safety invariants are that (i) a load cannot retire while required
po-older store-side ordering remains unresolved, and (ii) stores become
globally visible in ordering-tag order. Unsafe loads are blocked while
relevant lower-tag stores remain pending in either the store queue or the
merge buffer, and invalidation-matched loads are marked hazardous only when
an active ordering constraint exists (an incomplete po-older load-acquire
or pending lower-tag stores). At the same time, merge-buffer completion
preserves tag-ordered store visibility. Hazardous loads are then checked at
the corresponding ordering trigger, ensuring that no load is observed to
violate required ordering with po-older stores.

The key progress invariant is that retired lower-tag stores eventually leave
the store queue and drain from the merge buffer once their target cache lines
become available and they are eligible under the completion policy. As both
structures clear, waiting unsafe loads become retire-eligible and make progress.
Thus, \method need not couple ROB retirement to full merge-buffer drain in
order to ensure the required store visibility. This decoupling removes unnecessary
drain-induced retirement stalls yet preserves architectural ordering. All
required enforcement remains within-core in the base design, without adding
global coherence-protocol mechanisms.

\subsection{Closest Related Mechanism}
\label{sec:related}

Among prior proposals, zFence~\cite{aga:sin15} is the most directly
comparable to our approach. It reduces fence latency by decoupling
coherence-permission acquisition from data return for po-older stores and
allowing relaxed retirement once those stores are permission-ready and
protected. Building on this insight, the same early-permission condition can also relax
\textit{unsafe}-load retirement: an unsafe load can retire once
coherence permissions have been acquired for all relevant lower-tag pending
merge-buffer entries, even before those entries have drained.

\method targets a complementary bottleneck.  zFence accelerates retirement by making store-side permission readiness visible earlier, whereas \method decouples fence retirement from drain-based ordering by explicitly encoding ordering state in tags and enforcing it at retirement, merge-buffer
completion, and load-hazard checks.


To isolate mechanism contributions, we evaluate four designs:
baseline, zFence-only, \method-only, and \method+zFence
(Section~\ref{sec:analysis}). zFence-only captures the effect of early
permission and line-lock-protected relaxed retirement. In \method-only,
fence retirement is decoupled from coherence-permission readiness and can
proceed after the ordering-tag transition. The combined design removes both
bottlenecks: \method eliminates fence-head blocking, and zFence-style
permission checks further reduce \textit{unsafe}-load retirement waits.
In the combined design, permission requests are issued first for entries in
the current minimum-tag group, which preserves forward progress under
tag-ordered drain.

With these enforcement rules and comparison points established, we next
describe the simulation infrastructure and report results.

\section{Evaluation Methodology}
\label{sec:analysis}

\ignore{
This section evaluates whether \method preserves the targeted Armv8 ordering
semantics and quantifies where its performance benefit comes from. We first
describe the simulation infrastructure and correctness methodology, then
present the workloads and comparison points used throughout the evaluation.
}

\begin{table*}[t]
\centering
\footnotesize
\setlength{\tabcolsep}{3pt}
\caption{TEMPO features, RC rules, and an exhaustive Armv8 litmus suite at RC-rule granularity.}
\label{tab:litmus_feature_map}
\begingroup
\setlength{\arrayrulewidth}{0.3pt}
\begin{tabular}{@{}p{0.24\textwidth}!{\vrule width 0.3pt}p{0.34\textwidth}!{\vrule width 0.3pt}>{\raggedright\arraybackslash}p{0.24\textwidth}!{\vrule width 0.3pt}p{0.12\textwidth}@{}}
  \toprule
\multicolumn{1}{c!{\vrule width 0.3pt}}{Feature} &
\multicolumn{1}{c!{\vrule width 0.3pt}}{If missing or incorrect} &
\multicolumn{1}{c!{\vrule width 0.3pt}}{Litmus tests} &
\multicolumn{1}{c}{RC rule(s)} \\
\midrule
Full-fence tag transition and tag-ordered store completion &
Po-older stores can become globally visible after po-younger full-fence-separated accesses &
\texttt{SB+dmb.sy+rel-acq}, \texttt{SB+dmb.sy+rel-acqpc} &
\texttt{RC1} \\
  \midrule
Safe/unsafe load retirement predicate (unsafe-load completion is delayed until lower-tag store completion) &
Loads can retire before required po-older stores complete, violating ordering &
\texttt{CoWR}, \texttt{SB+dmb.sy+rel-acq}, \texttt{MP+rel+acq} &
\texttt{RC1, RC6} \\
  \midrule
Store-release higher-tag assignment and completion ordering &
A \texttt{store-release} can be observed before po-older regular stores &
\texttt{MP+rel+acq}, \texttt{MP+rel+acqpc}, \texttt{MP+rel+SWPacq} &
\texttt{RC3} \\
  \midrule
Load-acquire hazard tracking while the acquire is active &
Po-younger loads may commit stale values that should be ordered after the acquire &
\texttt{MP+rel+acq}, \texttt{MP+CAS-rfi-ctrl+acq} &
\texttt{RC2} \\
  \midrule
Read-after-read (RAR) hazard recovery for same-address load reordering &
Per-location ordering for same-address loads can be violated &
\texttt{CoRR} &
\texttt{RC5} \\
  \midrule
$RC_{SC}$ $RC4$ completion gating for load-acquire (delay completion until po-older \texttt{store-release} drain) &
\texttt{LDAR}/\texttt{STLR} sequential-ordering constraints can be violated &
\texttt{MP+rel+acq}, \texttt{MP+rel+swp-acq} &
\texttt{RC4} \\
  \bottomrule
\end{tabular}
\endgroup
\end{table*}

\subsection{Simulation Infrastructure}

We evaluate \method using the gem5 O3 CPU model~\cite{gem5,gem52020}. In
the default O3 model, post-fence memory accesses are serialized by making them
dependent on the fence, so they cannot issue before fence retirement. To
construct a realistic weak-memory baseline, we extend the core to support
speculative post-fence execution and enable it in all compared designs. We
also model an explicit merge buffer for post-retirement stores to capture the
retirement/completion behavior in Section~\ref{subsec:completion}.

To validate functional correctness, we run Armv8 litmus tests for the memory
model~\cite{alg:dea21,alg:mar14} before collecting performance results.
Table~\ref{tab:litmus_feature_map} maps each correctness-critical \method
feature to litmus tests that can fail if that feature is omitted or
implemented incorrectly. At the RC-rule level, the suite is exhaustive: each
rule RC1--RC6 is explicitly covered. SB-family tests primarily stress
store-side visibility ordering across fences, while MP-family tests stress
release/acquire ordering and load-hazard handling.

Our microarchitectural configuration is modeled after the Arm Neoverse V2
design point~\cite{bru23}. Table~\ref{tab:params} summarizes the simulated
parameters, including a cache hierarchy with a private 2~MiB L2 cache per core
and a shared 16~MiB L3 cache. All evaluated designs use a 320-entry ROB, 48
load/store issue-queue entries, a 32-entry load queue, a 64-entry store queue,
and a 16-entry merge buffer.

All experiments use gem5's syscall-emulation (SE) mode and the classic
memory hierarchy to keep the native multithreaded and DBT-proxy evaluation
feasible.  SE mode excludes OS and MMU side-effects; thus the reported
gains isolate user-level effects and are measured under the classic
snooping timing model shared by all compared designs.

\subsection{Hardware Cost}

Ordering tags use 10 bits, which is sufficient to cover the in-flight
window in our configuration. The added ordering-tag metadata in the
load/store issue queue, load queue, store queue, and merge buffer is
$(48 + 32 + 64 + 16)\times 10 = 1{,}600$ bits.

The ordering-tag queue has 16 entries (matching merge-buffer capacity). Each
entry stores a 10-bit tag, a 4-bit count, and a 1-bit fence marker, for
15 bits per entry and $16\times 15 = 240$ bits total. In addition, the
oldest-entry override uses a 16-bit snapshot vector per merge-buffer entry,
for $16\times 16 = 256$ bits. Overall, \method adds
$1{,}600 + 240 + 256 = 2{,}096$ bits (262 bytes) per core.
The ordering-tag queue maintains distinct active tags in modularly sorted
order, so the minimum is available at the head without a general associative
min-search. Allocations increment a matching entry or insert a new tag at its
sorted position; drains decrement the entry matching the drained store's tag.

On the store-to-load forwarding path, the match key becomes
\textit{(physical address, ordering tag)} rather than physical address
alone. For a 16-entry merge-buffer lookup, this widens the CAM comparison
by $16\times 10 = 160$ comparison bits relative to an address-only match. The
unsafe-load and hazard predicates also compare tagged store-queue entries,
adding logic but no storage. On the merge-buffer drain path, completion arbitration becomes a tag-gated
priority select: each entry compares its tag against the current minimum,
ANDs that result with cache-line readiness and oldest-entry-override
eligibility, and a 16-way priority encoder selects the drain candidate.

\subsection{Compared Designs}

We compare four designs: (i) baseline, (ii) zFence-only, (iii)
\method-only, and (iv) \method+zFence. All four designs use the same core,
cache, and memory parameters from Table~\ref{tab:params}; they differ only
in ordering handling.

The baseline does not use execute-time read-for-ownership (RFO) prefetching
for stores: a store issues its RFO prefetch request when it commits and enters
the merge buffer, rather than speculatively when its address is generated.
Adding speculative RFO prefetching at address generation would start ownership
acquisition earlier and reduce the exposed miss latency for stores whose
target lines can be obtained before commit, and would therefore reduce some
baseline drain stalls at fences and store-release operations. This would
likely narrow the measured benefit of \method in workloads where uncontended
store misses dominate the drain delay. The effect is not uniformly positive,
however: for contended cache lines, early RFOs can acquire ownership before the
store is ready to commit, increase invalidation traffic, and cause extra line
bouncing. Accordingly, the reported baseline does not incorporate execute-time
RFO prefetching.

zFence relaxes retirement only after acquiring
coherence permissions and line locks for the relevant merge-buffer entries.
\method instead retires fences after the tag transition and delays only
unsafe-load completion until lower-tag stores finish. \method+zFence keeps
that immediate fence retirement and additionally uses zFence permissions to
accelerate unsafe-load completion.

For zFence, we add per-line lock metadata in the L1 cache and a dedicated
lock-line request flag. Invalidating snoops to locked lines are deferred and
replayed after unlock; on a protection conflict, the core falls back to
drain-based retirement checks.

\subsection{Workloads}
\label{sec:workloads}

We evaluate two workload classes: native PARSEC~\cite{parsec}/
Rodinia~\cite{che:boy09} applications and a DBT-motivated SPEC proxy. All
benchmarks were
compiled with Clang 22.1, and the evaluated sets reflect the workloads we were
able to compile successfully for AArch64 and run to completion in gem5.

PARSEC and Rodinia represent native multithreaded execution, where ordering
instructions arise from synchronization in application code. We use PARSEC
\textit{simmedium}, Rodinia medium inputs, and four software threads. Native
statistics are reset at each region of interest (ROI) and collected through
its end; pre-ROI execution serves only as warmup. For native workloads, we report
performance as ROI speedup over the baseline, equivalently shown as
execution cycles normalized to baseline.

The second workload class targets cross-ISA DBT systems such as
QEMU~\cite{bel05}, where preserving guest TSO on an AArch64 host can greatly
increase fence frequency~\cite{lus:tri15,gao:men24}. Rather than model a full
DBT runtime, we use a controlled proxy: SPEC CPU2017
workloads~\cite{buc18spec} compiled with LLVM~\cite{lat04llvm} and
instrumented to insert a fence after each memory access, approximating a
conservative QEMU-style translation strategy. Real QEMU does perform some
fence fusion, but translated execution still exhibits similar fence density, so
the proxy remains a meaningful fence-dense stress case. SPEC runs use the
single-threaded rate configuration, restore SimPoint checkpoints, apply a
100M-instruction post-restore warmup, and then measure weighted
200M-instruction intervals. We report performance as IPC gain over the
baseline across the 200M-instruction measurement window.

\subsection{Reported Metrics}

We report performance, load/store queue full stalls, fence density, and
commit-side drain stalls. Native workloads use normalized execution cycles
because all runs share the same four-thread configuration and ROI boundaries;
this is equivalent to reporting ROI speedup over the baseline. The
instrumented SPEC proxy instead uses IPC because its single-threaded,
fence-dense setup makes IPC the clearest throughput signal.

\begin{table}[h!]
\centering
\footnotesize
\setlength{\tabcolsep}{3pt}
\caption{Simulated Architecture Parameters}
\begin{tabular}{@{}p{0.30\columnwidth}p{0.64\columnwidth}@{}}
  \toprule
Processor & 4-core, 3 GHz with 320-entry ROB \\
  \midrule
Width & 6-wide fetch/decode, 8-wide rename/issue/commit \\
  \midrule
Load/store ports & One load and two load/store ports \\
  \midrule
Load/store queues & 32-entry load queue / 64-entry store queue \\
  \midrule
Merge Buffer & 16-entry merge buffer \\
  \midrule
L1 Cache & 64 KB I-cache and D-cache per core \\
 & 4-cycle tag/data access latency \\
  \midrule
L2 Cache & 2 MiB private L2 cache per core \\
 & 11-cycle tag/data access latency \\
  \midrule
L3 Cache & 16 MiB shared L3 cache \\
 & 20-cycle tag/data access latency \\
  \midrule
Coherence & MOESI protocol \\
  \midrule
Memory & 2 GB DDR5-4400 (DDR5\_4400\_4x8 timing controller) \\
  \bottomrule
\end{tabular}
\label{tab:params}
\end{table}

\section{Results}

We first report baseline-versus-\method results, then use the native
PARSEC/Rodinia study to compare four configurations motivated by the
complementarity discussed in Section~\ref{sec:related}: baseline, zFence-only,
\method-only, and \method+zFence. The DBT proxy
SPEC study below, which models a fence-dense cross-ISA translation stress case,
remains a baseline-versus-\method comparison.

\subsection{Native Multithreaded Workloads}
\label{subsec:native_results}

\begin{figure}[t]
\includegraphics[width=\columnwidth]{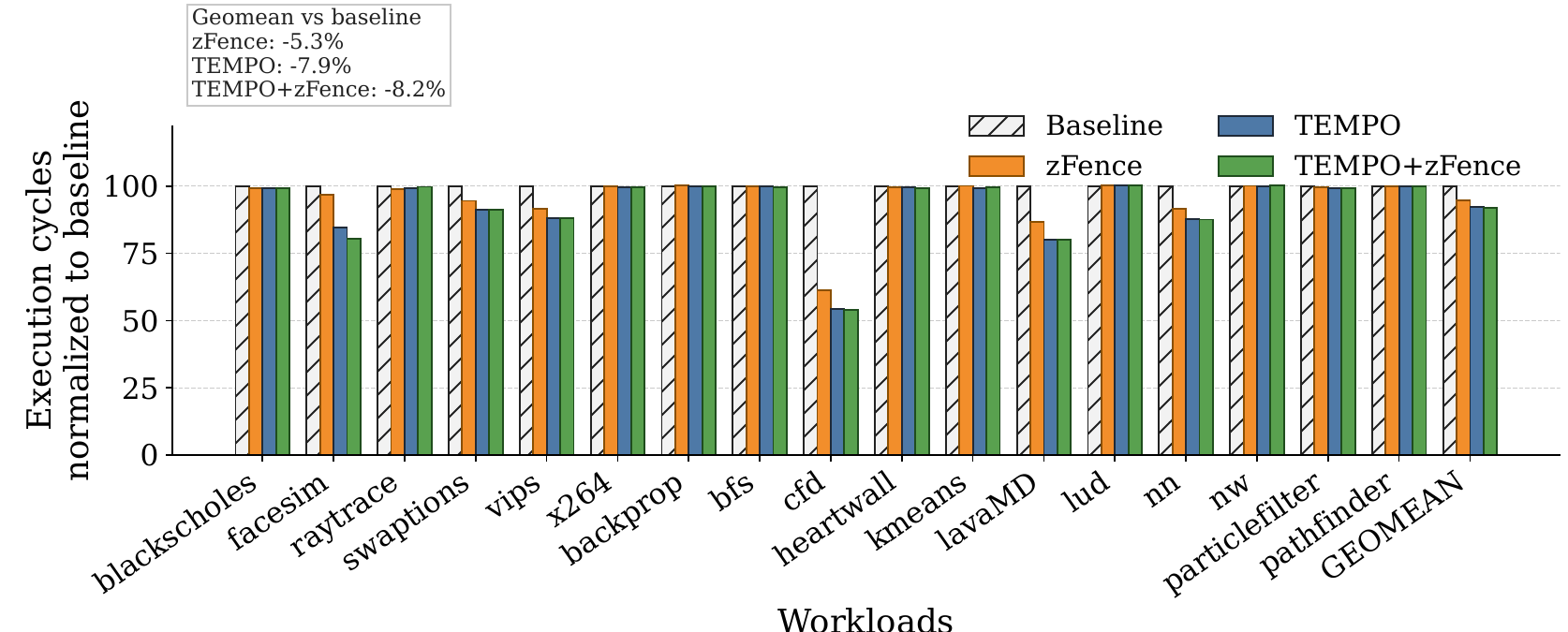}
\caption{Execution cycles normalized to baseline for native workloads
(4T) across baseline, zFence-only,
\method-only, and \method+zFence configurations.}
\label{fig:native_exec_cycles}
\end{figure}

Figure~\ref{fig:native_exec_cycles} reports execution cycles
normalized to the baseline for the native four-thread workloads. \method-only
reduces geometric-mean execution cycles by $7.9\%$, compared with $5.3\%$ for
zFence-only; the combined design reaches $8.2\%$, only $0.3$ percentage points beyond
\method alone. Thus, ordering tags capture most of the recoverable native
benefit, while zFence adds mainly when po-older stores are tied to
long-latency misses. In particular, \method delivers broad gains across
\texttt{facesim}, \texttt{swaptions}, \texttt{vips}, \texttt{nn}, and the
large \texttt{cfd}/\texttt{lavaMD} cases, whereas zFence's strongest
standalone gains remain concentrated in \texttt{cfd} and \texttt{lavaMD}.

\begin{figure}[t]
\includegraphics[width=\columnwidth]{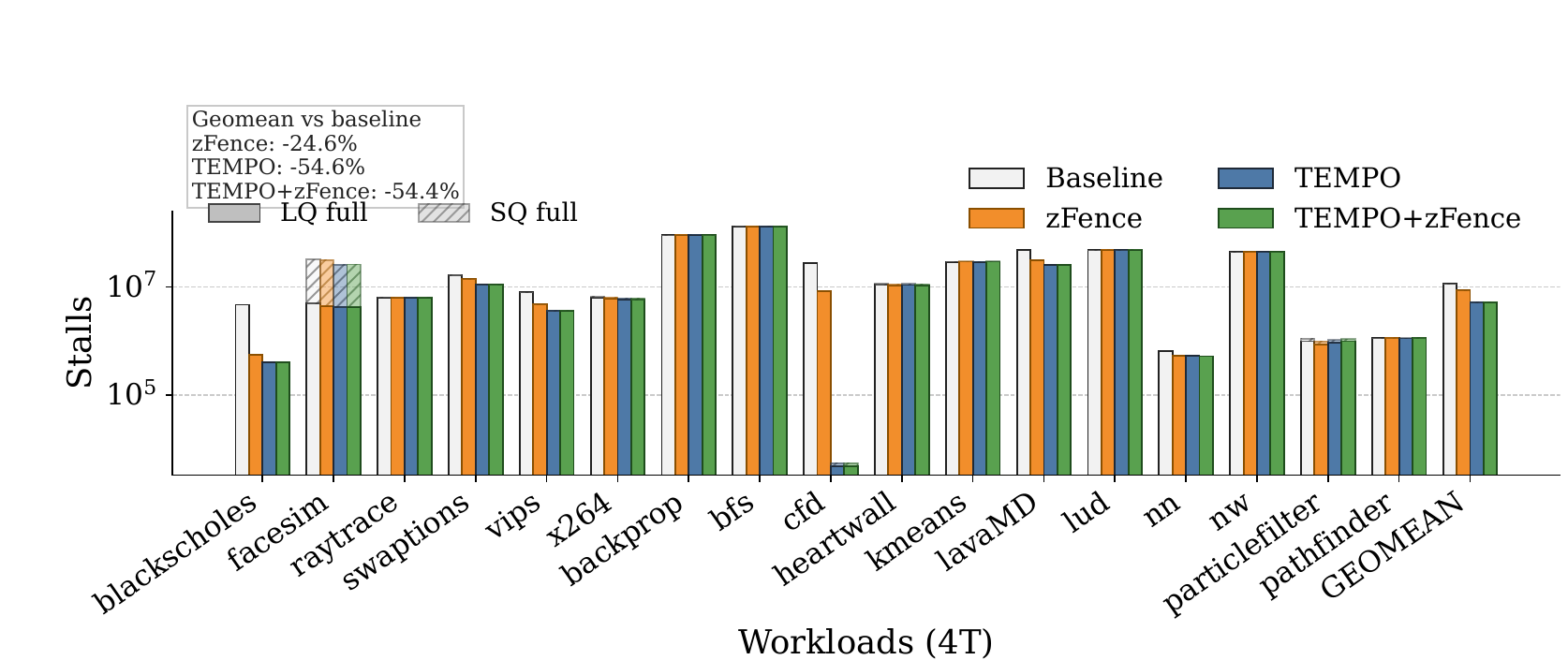}
\caption{Load/store queue full stalls in native workloads (4T),
aggregated across LQ-full and SQ-full stall counts. The y-axis uses a log scale.}
\label{fig:native_lqsq_full}
\end{figure}

Figure~\ref{fig:native_lqsq_full} shows combined LQ-full and SQ-full
stall counts: zFence-only reduces the geometric mean by $24.0\%$, while \method and
the combined design reduce it by $54.5\%$ and $54.2\%$, respectively. The
combined design adds little beyond \method, reinforcing that most native
benefit comes from load-side backpressure relief. Near-neutral workloads have
modest ordering pressure or load-queue bottlenecks; \texttt{raytrace} has more
ordering instructions, but negligible baseline drain-stall cost. In
\texttt{bfs} and \texttt{heartwall}, queue-full stalls barely move because
load-queue pressure, not ordering-induced blocking, dominates. More broadly,
most near-neutral native workloads have very low ordering-instruction density,
so there is little retirement-time ordering pressure to remove.

\begin{figure}[t]
\includegraphics[width=\columnwidth]{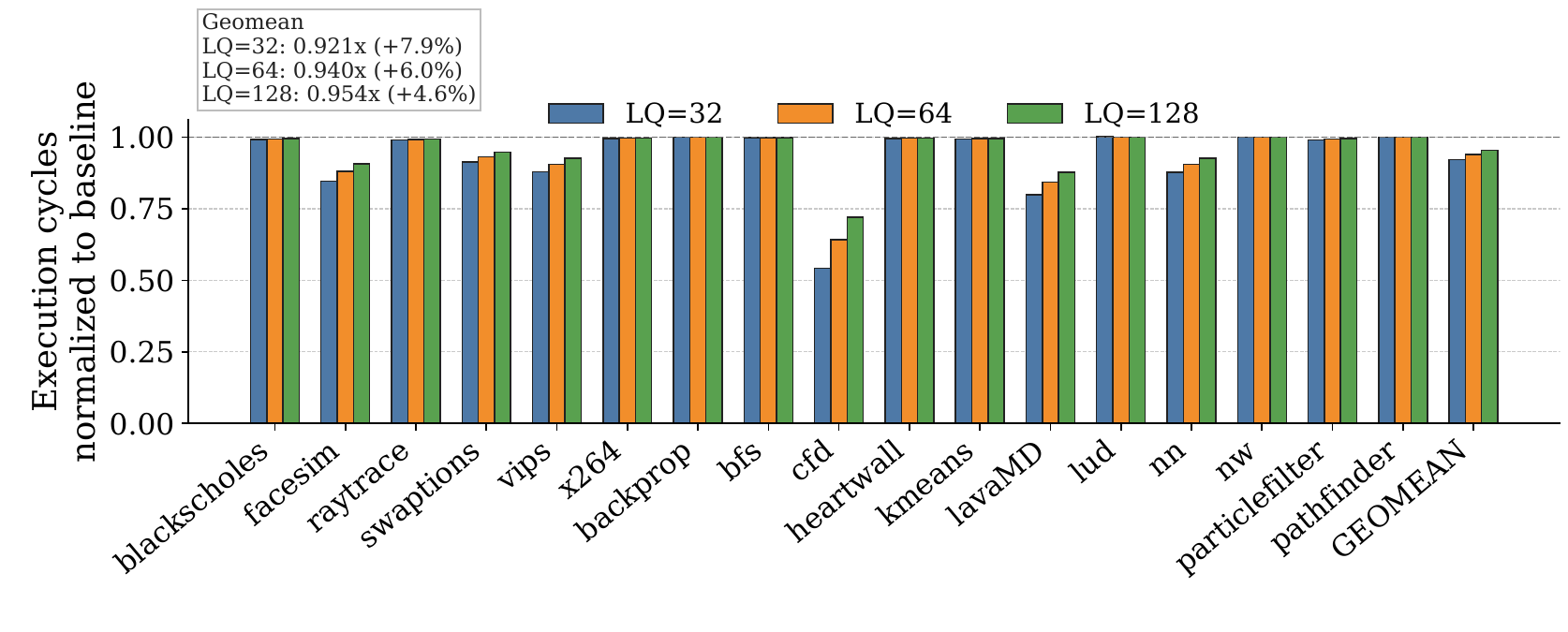}
\caption{Execution cycles normalized to baseline for each native
  workload at 32-, 64-, and 128-entry load-queue capacities; lower is
  better.}
\label{fig:native_lq_sensitivity}
\end{figure}

Figure~\ref{fig:native_lq_sensitivity} shows that \method's native gains fall
from $7.9\%$ at 32 LQ entries to $4.6\%$ at 128, while the LQ-sensitive subset
still gains $11.3\%$ at 128 entries. This trend is consistent with
Figure~\ref{fig:native_lqsq_full}: \method's gains are larger when smaller
load queues make conservative ordering pressure visible as execution-time loss.

\subsection{Instrumented SPEC2017 (DBT Proxy)}

\subsubsection{Performance and Fence Intensity}

\begin{figure}[t]
\includegraphics[width=\columnwidth]{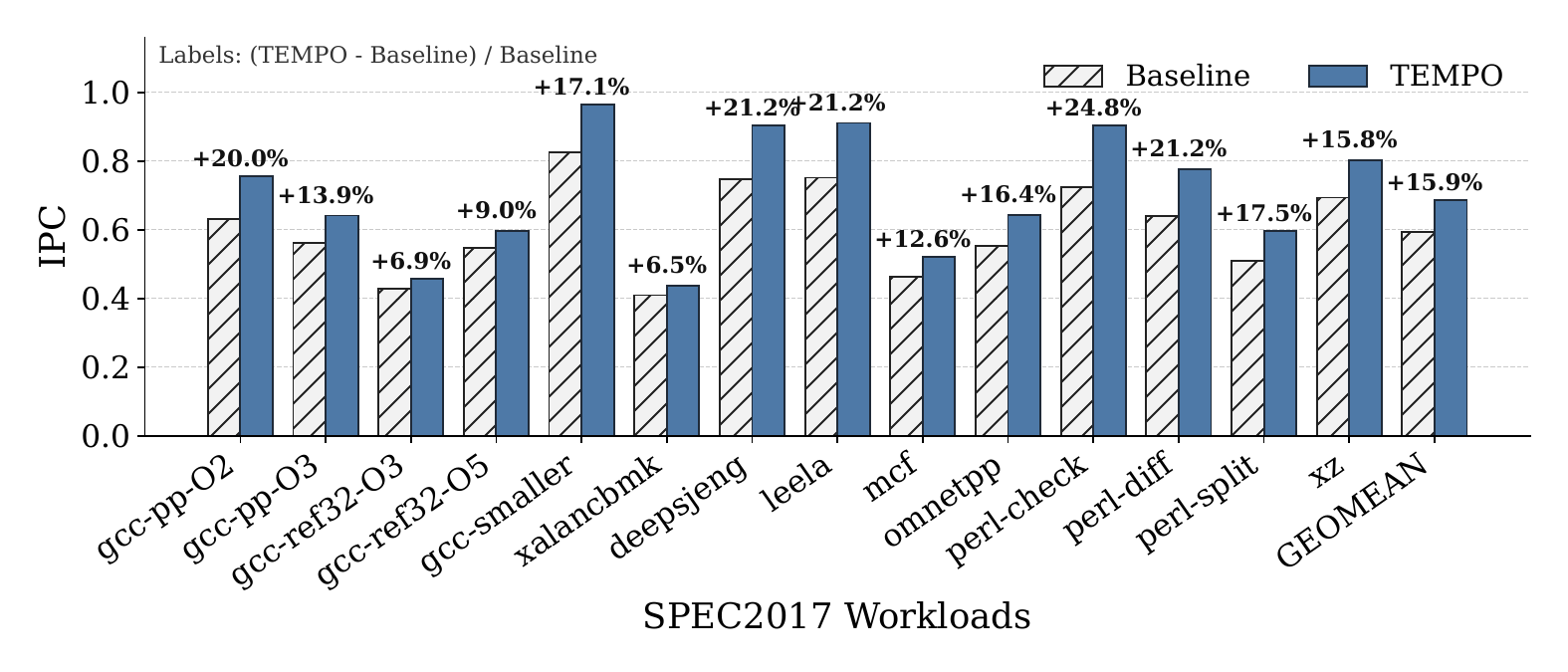}
\caption{IPC in baseline vs.\ \method for instrumented SPEC2017.}
\label{fig:spec_ipc}
\end{figure}

\begin{figure}[t]
\includegraphics[width=\columnwidth]{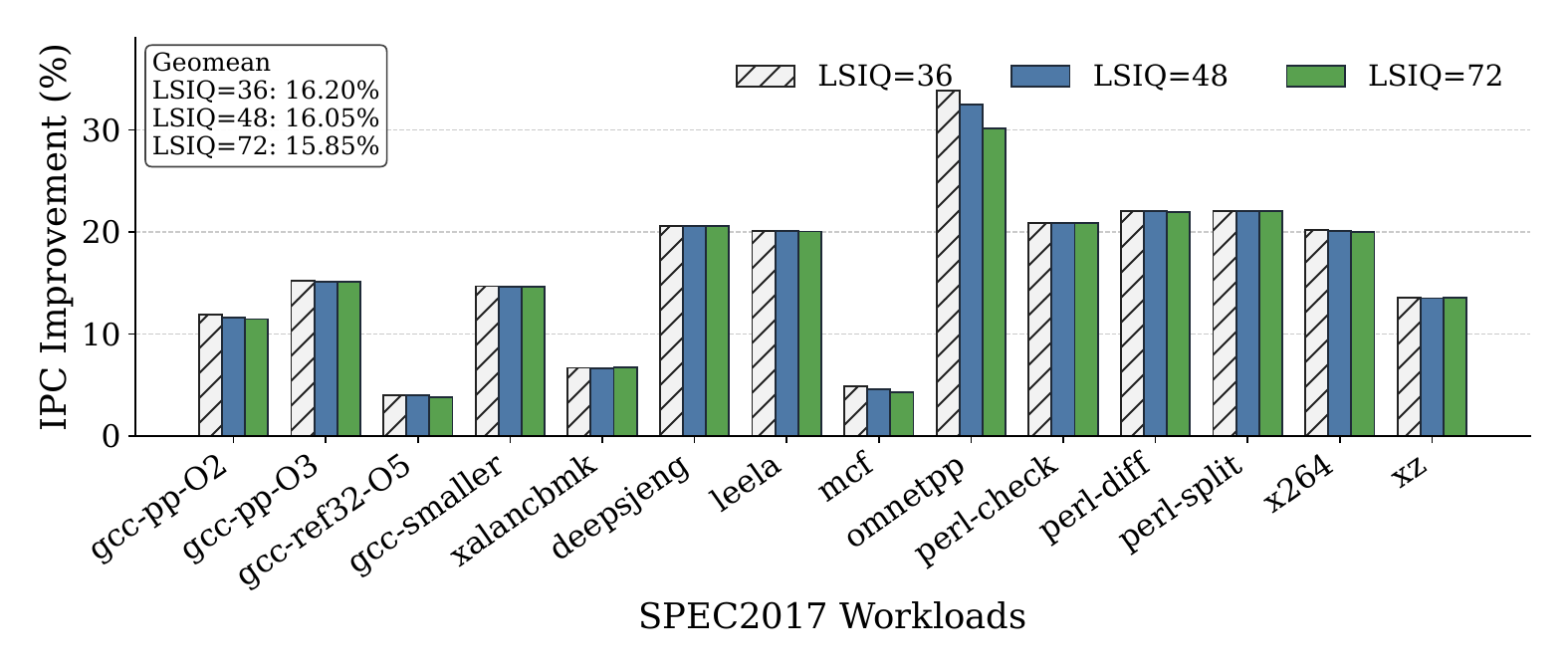}
\caption{IPC improvement of \method over baseline for SPEC2017 LSIQ sweeps
with 128-entry LQ and 64-entry SQ.}
\label{fig:spec_lsiq_sweep}
\end{figure}

Figure~\ref{fig:spec_ipc} reports IPC for the instrumented SPEC CPU2017
workloads. \method improves geometric-mean IPC by $15.9\%$ over the
baseline. This workload serves as a proxy for cross-ISA DBT execution, such as
x86-on-Arm, where conservative translation can inject many fences. It is
therefore deliberately fence-dense, so performance is heavily shaped by
ordering overhead. Figure~\ref{fig:spec_lsiq_sweep} shows that geometric-mean
gains remain stable as LSIQ capacity increases from 36 to 72 entries
($16.2\%$ and $15.9\%$, respectively).

The average fence density is $187$ fences per thousand instructions.  At
this rate, reducing retirement-time fence stalls has a first-order impact on
throughput. Even moderate reductions in fence residency and commit waiting
therefore translate into visible IPC gains.

\subsubsection{Retirement and Commit Behavior}

\begin{figure}[t]
\includegraphics[width=\columnwidth]{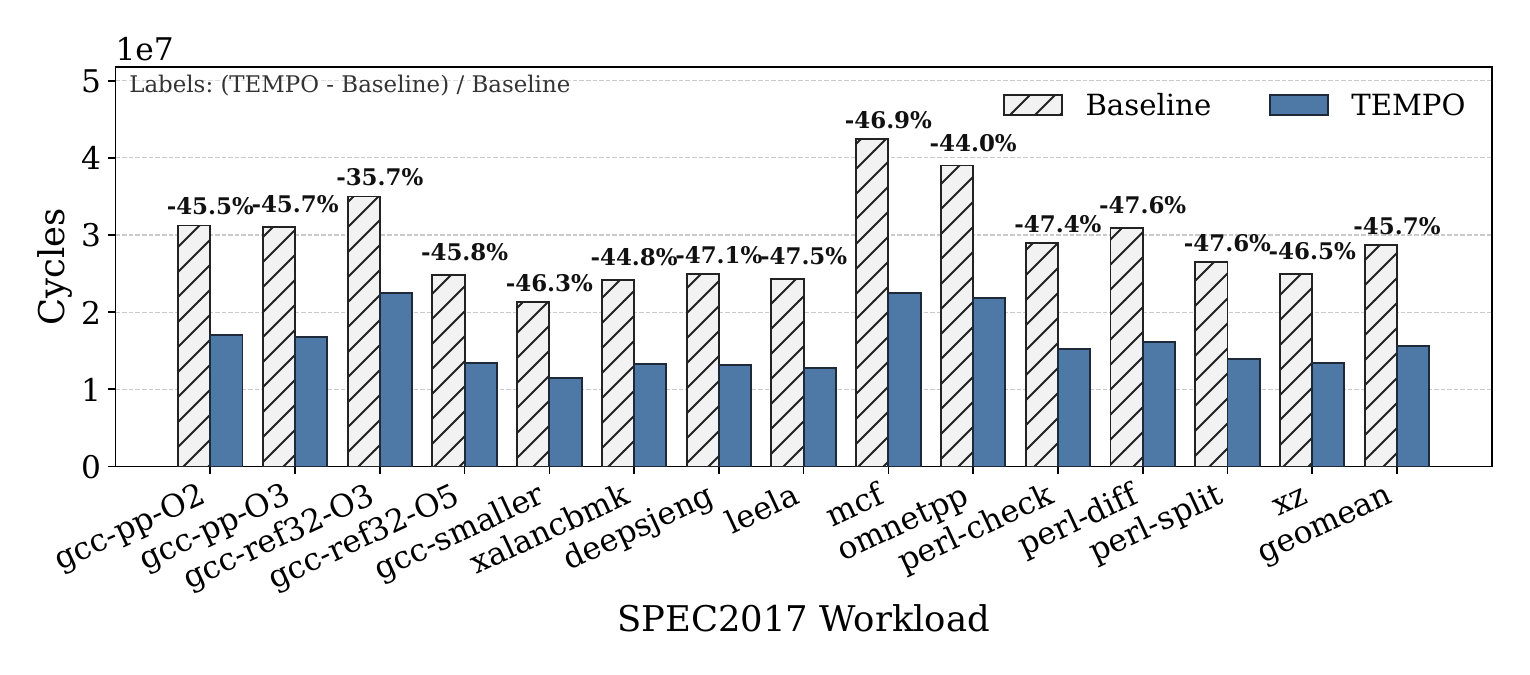}
\caption{Commit stall cycles waiting for merge-buffer drain.}
\label{fig:commit_stalls}
\end{figure}

Figure~\ref{fig:commit_stalls} shows reduced commit stalls. The
baseline's fence-wait stalls are largely eliminated and partly
redistributed to \textit{unsafe}-load retirement waits. This shift is
expected: \method replaces conservative fence-wide stalls with finer-grain
ordering checks at load retirement. Even so, total commit-side waiting is
reduced, contributing to the IPC improvement in Figure~\ref{fig:spec_ipc}. In
particular, commit stalls waiting for merge-buffer drain fall by $45.7\%$ on
the geometric mean. This does not translate one-to-one into IPC because only the
critical-path portion of these stalls is recoverable, while other overlapping
backend bottlenecks still limit throughput.

\ignore{

\subsubsection{Merge-Buffer Pressure}

Figure~\ref{fig:mb_occupancy} shows an increase in merge-buffer occupancy in
\method. This is a direct consequence of earlier fence retirement:
stores enter the merge buffer sooner, increasing in-flight post-retirement
store state. The occupancy increase is therefore an expected side effect of
decoupling retirement from completion. Despite this higher occupancy,
throughput improves (Figure~\ref{fig:spec_ipc}), indicating that reduced
retirement stalls outweigh additional merge-buffer pressure in these
workloads.

\begin{figure}[t]
\includegraphics[width=0.5\textwidth]{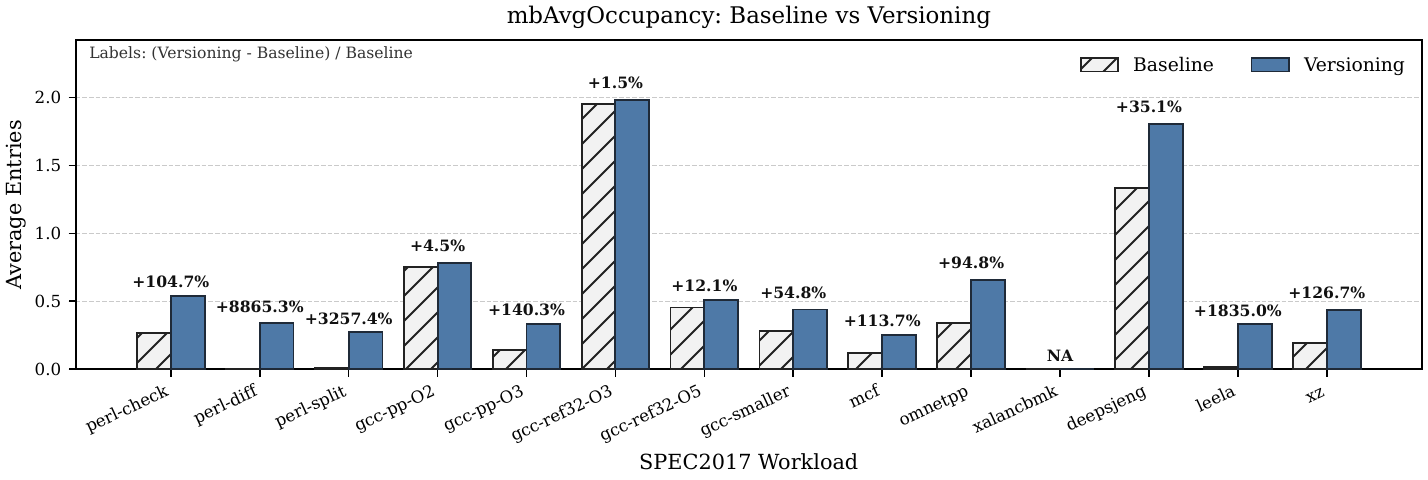}
\caption{Merge-buffer occupancy in instrumented SPEC2017 workloads.}
\label{fig:mb_occupancy}
\end{figure}
}

\section{Discussion: Applicability to TSO}
\label{sec:discussion}


\method is primarily targeted at weak-memory architectures such as Armv8 and
RISC-V, where explicit ordering instructions interact with an unordered
post-retirement store path. A TSO-style adaptation is possible
in principle~\cite{sew:sar10}, but its benefit is narrower. Under TSO, store
global visibility already follows FIFO completion in the store-buffer/merge-buffer
path, so most store-side ordering is already enforced by the baseline
microarchitecture. The main remaining opportunity is therefore decoupling
retirement of strongly ordered instructions such as \texttt{mfence} from store
drain, rather than exploiting tag-ordered completion as broadly as in the RC
setting. We therefore view TSO as a possible extension path, not the primary
use case for \method.

\section{Additional Related Work}

Beyond zFence, prior work on reducing fence overhead falls into three
categories: software translation/optimization, ISA-assisted ordering, and
speculation- or coherence-assisted hardware. These categories differ in where
they intervene: before execution (software), at the ISA contract (new ordering
primitives), or at runtime (microarchitecture/coherence).

Software approaches reduce ordering cost before execution by transforming the
fence stream (inference, elimination, weakening, or application-level
rewrites)~\cite{kup:vec10,vaf:nar11,liu:zan20}. In cross-ISA settings, Risotto
and Lasagne target strong-on-weak translation through verified mapping rules
and optimized fence insertion in dynamic and static translators,
respectively~\cite{gou:spr22,roc:spr22}. These approaches reduce fence
frequency, but once fences remain in the instruction stream, they do not
directly remove retirement-time bottlenecks in the core pipeline.

ISA-assisted mechanisms reduce conservative enforcement by exposing finer-grain
ordering semantics, including address-aware and scoped/asymmetric fence forms,
as well as dependence-directed ordering (EDE)~\cite{lin:nag13,lin:nag14,dua:hon15,shu:vou21}.
Address-aware fences limit ordering to selected addresses, scope-aware fences
limit ordering to a visibility domain (e.g., local vs. system-wide), and
asymmetric forms encode directional constraints (e.g., load-facing or
store-facing order). Related proposals also classify accesses (e.g., shared
versus local) to apply differentiated buffering and avoid unnecessary
serialization~\cite{sin:nar12}.
These methods shrink the scope of enforced order through ISA/policy changes,
whereas \method keeps the ISA unchanged and refines enforcement
microarchitecturally.

A third line of work uses speculation and/or coherence-visible machinery to
preserve correctness while relaxing conservative stalls. WeeFence and
InvisiFence rely on conflict tracking and rollback-aware speculative
execution~\cite{dua:muz13,blu:mar09}. Free Atomics and bounded atomic-region
retry techniques reduce fence-like serialization around atomic
operations~\cite{asg:ceb22,gom:edu25}. Non-Speculative Load-Load Reordering in
TSO avoids unnecessary load squash/replay by hiding observed load reordering
through protocol-level handling~\cite{ros:car17}.

Compared with these mechanisms, \method targets the same high-level goal
(removing unnecessary serialization and replay) with a different deployment
point: in its base form, it is a within-core mechanism that does not
require ISA extensions or coherence-protocol modifications. As shown in
Section~\ref{subsec:native_results}, this design is also complementary to
zFence-style permission-based acceleration.

\section{Conclusion}

We introduce the concept of ordering tags, which \method uses to enforce only
the required memory-ordering edges at retirement and completion, avoiding broad
fence-induced serialization and unnecessary load recovery.
At retirement, unsafe loads wait only while lower-tag stores remain pending in
the store queue or merge buffer. At completion, the sorted ordering-tag queue
controls store visibility while the oldest-entry override preserves progress.
Across native
multithreaded workloads, \method reduces
geometric-mean normalized execution cycles by 7.9\% and load/store queue full
stalls by 54.6\%; \method+zFence reduces the latter by 54.4\%. In the
instrumented SPEC2017 DBT proxy, \method improves
geometric-mean IPC by 15.9\% and reduces commit stalls waiting for merge-buffer
drain by 45.7\%. The design adds only 262 bytes of metadata per core. The
central insight is that ordering need not imply pipeline-wide serialization.
Explicit tags preserve required edges while allowing unrelated accesses to
progress. The
framework requires no ISA extensions or coherence-protocol changes.

\ignore{
\section*{Acknowledgment}
The authors thank Robert Clancy and Anoop Iyer at Arm for their
valuable feedback, which helped improve the clarity and
presentation of this work.  This research was supported in part by the
National Science Foundation (NSF) under Grant SHF-2452082.
}

\clearpage

\bibliographystyle{IEEEtran}
\bibliography{ref}


\end{document}